\documentstyle[aaspp4m,psfig,11pt]{article}

\def\unsetyr{\def\oyear{\relax}\def\cyear{\relax}\def\cyeara{a\relax}\def\cyearb{b\relax}\def\cyearc{c\relax}\def\cyeard{d\relax}}
\def\setyr{\def\oyear{(}\def\cyear{)}\def\cyeara{a)}\def\cyearb{b)}\def\cyearc{c)}\def\cyeard{d)}}
\unsetyr
\def\jcite#1{\setyr\cite{#1}\unsetyr}

\def\rmmat#1{{\hbox{\rm #1}}}
\def\rmscr#1{\rmmat{\scriptsize #1}}
\newcommand{\be}{\begin{equation}}
\newcommand{\ee}{\end{equation}}
\newcommand{\bt}{\begin{table} \begin{center}}
\newcommand{\et}{\end{center} \end{table}}
\newcommand{\ba}{\begin{eqnarray}}
\newcommand{\ea}{\end{eqnarray}}
\newcommand{\ie}{{\it i.e.~}}
\newcommand{\eg}{{\it e.g.~}}
\def\p{\partial}
\def\d{{\rm d}}

\def\dd#1#2{\frac{\d #1}{\d #2}}
\def\pp#1#2{\frac{\p #1}{\p #2}}
\newcommand{\comment}[1]{\relax}
\def\eqref#1{Equation~\ref{eq:#1}}

\def\figref#1{Figure~\ref{fig:#1}}

\begin{document}
\newcommand{\bfi}{{\bf B}} \newcommand{\efi}{{\bf E}}
\newcommand{\lel}{{\lambda_e^{\!\!\!\!-}}}
\newcommand{\me}{m_e}
\newcommand{\mcs}{{m_e c^2}}
\title{Polarization Evolution in Strong Magnetic Fields}
\author{Jeremy S. Heyl}
\authoremail{jsheyl@tapir.caltech.edu}
\and
 \author{Nir J. Shaviv,}
\authoremail{nir@tapir.caltech.edu}
\affil{Theoretical Astrophysics, mail code 130-33,
California Institute of Technology, Pasadena, CA 91125}
\begin{abstract}
Extremely strong magnetic fields change the vacuum index of
refraction.  Although this polarization dependent effect is small for
typical neutron stars, it is large enough to decouple the polarization
states of photons traveling within the field.  The photon
states evolve adiabatically and follow the changing magnetic field
direction. The combination of a rotating magnetosphere and a frequency
dependent state decoupling predicts polarization phase lags between
different wave bands, if the emission process takes place well within
the light cylinder. This QED effect may allow observations to
distinguish between different pulsar emission mechanisms and to
reconstruct the structure of the magnetosphere.
\end{abstract}

\section{Introduction}

Understanding the structure of the magnetic fields surrounding neutron
stars may provide a key in developing models for the radio and X-ray
emission from pulsars, pulsar spindown, soft-gamma repeaters and the
generation of the magnetic fields themselves.  Although the magnetic
field is instrumental in models of many phenomena associated with
neutron stars, measuring its structure over a range of radii is
problematic.  Observations of the thermal emission from the surface
may constrain the magnetic field geometry near the star, and the
slowing of the pulsar's rotation may yield an estimate of the strength
of the field near the speed-of-light cylinder.  Connecting these
regimes is difficult.

The intense magnetic fields associated with neutron stars 
influence many physical processes -- cooling (\cite{Heyl97magnetar};
\cite{Shib95}), atmospheric emission (\cite{Raja97}; \cite{Pavl94})
and the insulation of the core (\cite{Heyl98numens};
\cite{Heyl97analns}; \cite{Scha90a}; \cite{Hern85}).  Even stronger
fields such as thought to be found near AXPs and SGRs alter the
propagation of light through the magnetosphere by way of
quantum-electrodynamic (QED) processes and may further process the
emergent radiation (\cite{Heyl97hesplit}; \cite{Heyl97index};
\cite{Bari95a}; \cite{Bari95c}; \cite{Adle71}), and distort our view
of the neutron star surface (\cite{Shav98lens}).

Even for weaker fields, QED renders the vacuum anisotropic.  The speed
at which light travels through the vacuum depends on its direction,
polarization and the local strength of the magnetic field.  Although
for neutron stars with $B < 10^{14}$~G this effect is too weak to
grossly affect images and light curves of these objects, it is strong
enough to decouple the propagating modes through the pulsar
magnetosphere.  

To lowest order in the ratio of the photon energy to
the electron rest mass-energy, the index of refraction of a photon in a
magnetic field is independent of frequency.   Near the pair-production 
threshold, the photon propagation adiabatically merges with a postironium 
state (\cite{Shab86}).  However, well above the pair-production 
threshold for weak fields ($m_e c^2 \ll E \ll m_e c^2 4.4 \times
10^{13}~\rmmat{G}/B$) the low 
energy results again provide a good approximation (\cite{Tsai75})
In the context of the field surrounding a neutron star, only photons
with sufficiently high wavenumbers (the optical and blueward) will
travel through a portion of the rapidly weakening magnetic field
without their two polarization modes mixing.

At radio frequencies, the plasma surrounding the neutron star produces
a similar effect (\eg \cite{Chen79}, \cite{Barn86}).  One would expect
that radio emission will be initially polarized according to the
direction of the local magnetic field.  When one observes a pulsar at
a particular instant, one sees emission from regions with various
magnetic field directions; therefore, one would expect the
polarization to cancel out substantially.  However, one finds that pulsars
exhibit a significant linear polarization.  As the polarized radiation
travels from its source, its polarization direction changes as the
local magnetic field direction changes.  At a distance from the star
that is large when compared to its radius, the local magnetic field in
the plane of the sky is parallel across the observed portion of the
star.  Since the field changes gradually, the polarization modes
are decoupled, and the disparate linear polarizations can add coherently.

Previous authors have focussed on the propagation of polarized radio
waves through the magnetosphere.  At these frequencies, the vacuum
polarization is safely neglected.  Furthermore, they have assumed that
the coupling of the two polarization modes occurs instantaneously.  In
this paper, we will treat the problem of vacuum polarization in
particular and how the gradual coupling of the polarization modes
affects the final polarization of the emergent radiation.  At
sufficiently high frequency the plasma only negligibly affects the
radiation as it travels through the magnetosphere.  The precise
frequency at which the vacuum birefringence begins to dominate depends
on the charge density of the magnetospheric plamsa.  If we assume the 
Goldreich-Julian value (\cite{Gold69}), we find that for
\be
E_\rmscr{photon} > 0.035\,\rmmat{eV}\, \left
( \frac{B}{10^{12}~\rmmat{G}}
  \frac{P}{1~\rmmat{sec}} \frac{n_\rmscr{GJ}}{n_e} \right )^{-1/2},
\ee
the vacuum contribution to the birefringence dominates that of
the plasma (the modes collapse where equality obtains (\cite{Mesz92})).
Our study of the vacuum-dominated regime, the optical and blueward, 
complements the work of \jcite{Chen79} and \jcite{Barn86} and will help to 
interpret observations of high-energy polarized radiation from 
neutron stars.

\section{Small Amplitude Waves in the QED Vacuum}

\jcite{Kubo81} (also \cite{Kubo83}) derive the equation of motion of
polarization direciton on the Poincar\'e sphere as light travels
through an inhomogeneous birefringent medium.  Since their results
assume that the medium is polarized but not magnetized, we first
extend their results to include magnetization.  The general equations
derived describe how polarized radiation travels through any
birefringent medium in the limit of geometric optics.  We then focus
on the propagation of high frequency radiation through pulsar
magnetospheres and how measurements of the polarization of this
radiation can constrain both the structure of the magnetic field and
the emission process of the radiation.  This extended formalism is
more than adequate to also describe the plasma induced birefringence
of radio waves. This could be used to extend the works of
\jcite{Chen79} and \jcite{Barn86}.

\jcite{Kubo83} find that the evolution of the polarization of a wave 
traveling through a birefringent and dichroic medium in the limit
of geometric optics is given by
\newcommand{\hatom}{{\bf {\hat \Omega}}}
\be
\pp{\bf s}{x_3} = \hatom \times {\bf s} + 
\left ( {\bf {\hat T}} \times {\bf s} \right ) \times {\bf s},
\label{eq:sevol}
\ee
where $x_3$ is the direction of propagation, ${\bf s}$ is the
normalized Stokes vector (\cite{Jack75}), and $\hatom$ and
${\bf {\hat T}}$ are the birefringent and dichroic vectors.  
The Stokes vector consists of the four Stokes parameters, $S_0, S_1,
S_2$ and $S_3$.  The vector ${\bf s}$ consists of $S_1/S_0, S_2/S_0$
and $S_3/S_0$. 

Waves traveling through a magnetized (or electrified) vacuum are best
treated by linearizing the constitutive relations about the external
field.  \jcite{Heyl97index} find that the permeability and dielectric
tensors consist of an isotropic component plus an added component along
the direction of the external field.
Using the dielectric and permeability tensors of a magnetized vacuum 
in the formalism of 
\jcite{Kubo83} yields
\be
 \hatom = 
\frac{k_0}{2 \sqrt {\mu_i \varepsilon_i + 1/2 \left ( \mu_i
\varepsilon_f + \mu_f \varepsilon_i \right ) \sin^2\theta}}
 \left (  \mu_i \varepsilon_f - \mu_f \varepsilon_i 
\right ) \sin^2\theta \left [ \begin{array}{c}
\phantom{-}\cos 2\phi  \\
-\sin 2\phi \\
0
\end{array}
\right ].
\label{eq:omegaqeddef}
\ee
where $\theta$ is the angle between the direction of propagation (${\bf
k}$) and the external field and $\phi$ is the angle between the
component of the external field perpendicular to ${\bf
k}$ and the $x-$axis defined by the observer.  $\mu_i$ and $\mu_f$ are the
isotropic and along the field components of the permeability tensor.
$\varepsilon_{i}$ and $\varepsilon_{f}$ are the equivalent dielectric
tensors.

For the case of vacuum QED with an external magnetic field to one loop,
we find that the amplitude of $\hatom$ is proportional to the difference 
between the indices of refraction for the two polarization states:
\ba
 {\Omega / k_0}  &=& \Delta n = n_\perp - n_\| \\ &=& \frac{\alpha}{4\pi}
 \sin^2 \theta \left [ -X_0^{(2)}\left(\frac{1}{\xi}\right) \xi^{-2} +
 X_0^{(1)} \left(\frac{1}{\xi}\right) \xi^{-1} - X_1
 \left(\frac{1}{\xi}\right) \right ]
\ea
to lowest order in $\alpha$, the fine-structure constant.
$\xi=B/B_\rmscr{QED}$ ($B_\rmscr{QED} \approx 4.4 \times 10^{13}$~G)
and the functions $X_0(x)$ and $X_1(x)$ are defined in
\jcite{Heyl97hesplit}.  An external electric field yields similar
results.  However, in this case the vacuum is also dichroic so the
vector ${\bf {\hat T}}$ is nonzero.

In the weak magnetic field limit ($\xi \ll 0.5$) we obtain,
\be
n_\perp - n_\| = \frac{\alpha}{4\pi} \frac{2}{15} \xi^2 \sin^2\theta,
\label{eq:delta_n_weak}
\ee
and the strong field limit ($\xi \gg 0.5$) yields
\be
n_\perp - n_\| = \frac{\alpha}{4\pi} \frac{2}{3} \xi \sin^2\theta.
\ee

\subsection{Exact Solutions}

\jcite{Kubo83} found that \eqref{sevol} yields exact solutions for
restricted values of $\hatom$ and ${\bf {\hat T}}$.  In the case of a
uniformly magnetized vacuum, ${\bf {\hat T}}=0$ and ${\bf {\hat
\Omega}}$ is constant.  In this case, the polarization vector ${\bf
s}$ traces a circle on the Poincar\'e sphere about the vector $\hatom$
at a rate of $|\hatom|$.

\jcite{Kubo81} examine the case where $\hatom$ is constant 
in magnitude but satisfies,
\be
\pp\hatom{x_3} = {\bf \Upsilon} \times \hatom
\ee
where ${\bf \Upsilon}$ is a constant.  In this case $\hatom$ rotates
about ${\bf \Upsilon}$ at a rate of $|{\bf \Upsilon}|$.  If we follow
the equations in a rotating coordinate system such that in it
$\hatom'$ is constant, we find that ${\bf s}'$ satifies the following
equation
\be
\pp{{\bf s}'}{x_3} = \hatom_\rmscr{eff} \times {\bf s}'
\label{eq:sevolrot}
\ee
where
\be
\hatom'_\rmscr{eff} = \hatom' - {\bf \Upsilon}.
\label{eq:omegaeff}
\ee
This equation holds even if ${\bf \Upsilon}$ is not constant.
However, if ${\bf \Upsilon}$ is constant, we obtain a new exact
solution where ${\bf s}$ circles a guiding center displaced from
$\hatom$ which in turn rotates about ${\bf \Upsilon}$.

If we take ${\bf s}~ \|~ \hatom$ initially and ${\bf \Upsilon} \perp
\hatom$, we find that ${\bf s}$ develops a component perpendicular to
$\hatom$.  In the case of vacuum QED, $\hatom$ lies in the $1-2-$plane.  If
the magnetic field rotates uniformly in the plane transverse to the
wave, we find that ${\bf s}$ will leave the $1-2-$plane and a
circularly polarized component will develop.

\subsection{Adiabatic Approximation}
\label{sec:adiabatic}

If the parameters describing the motion of a system vary slowly
compared to the characteristic frequency of the system, the system
evolves adiabatically such that at a given time it executes a motion
given by the instantaneous values of the parameters as if they were
static.  The exact solution given by \eqref{sevolrot} has this feature
if $|\hatom| \gg |{\bf \Upsilon}|$.  If this limit applies,
$\hatom_\rmscr{eff}$ is nearly parallel to $\hatom'$ and ${\bf s}$
circles the instantaneous guiding center $\hatom$.  Furthermore, if
${\bf s}$ is initially parallel or antiparallel to $\hatom$, it will
remain so (corresponding to polarizations that are parallel or
perpendicular to the direction of the magnetic field).  That is, the
polarization modes are decoupled, and the polarization direction
follows the direction of the magnetic field.

\begin{figure}
\plottwo{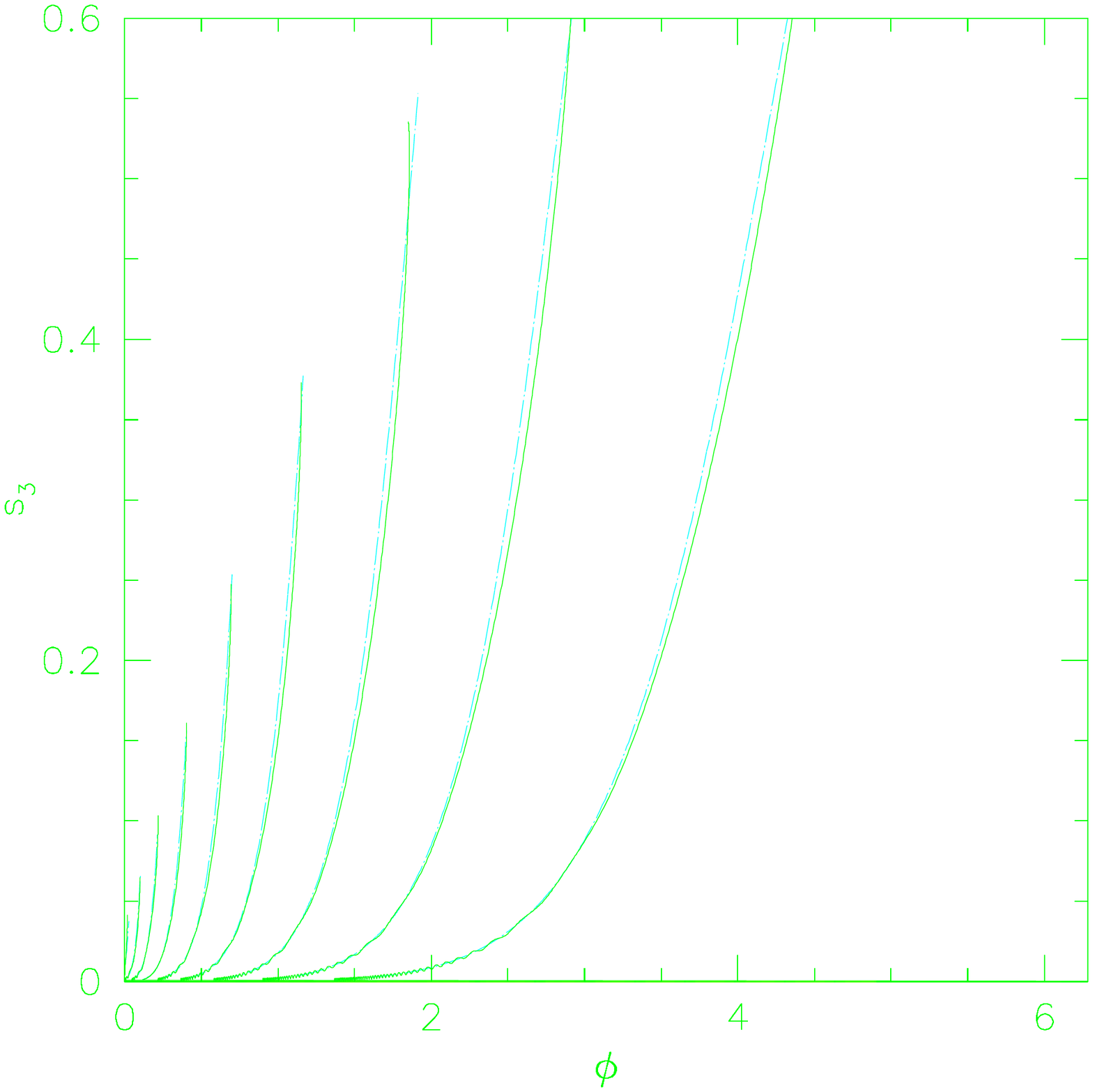}{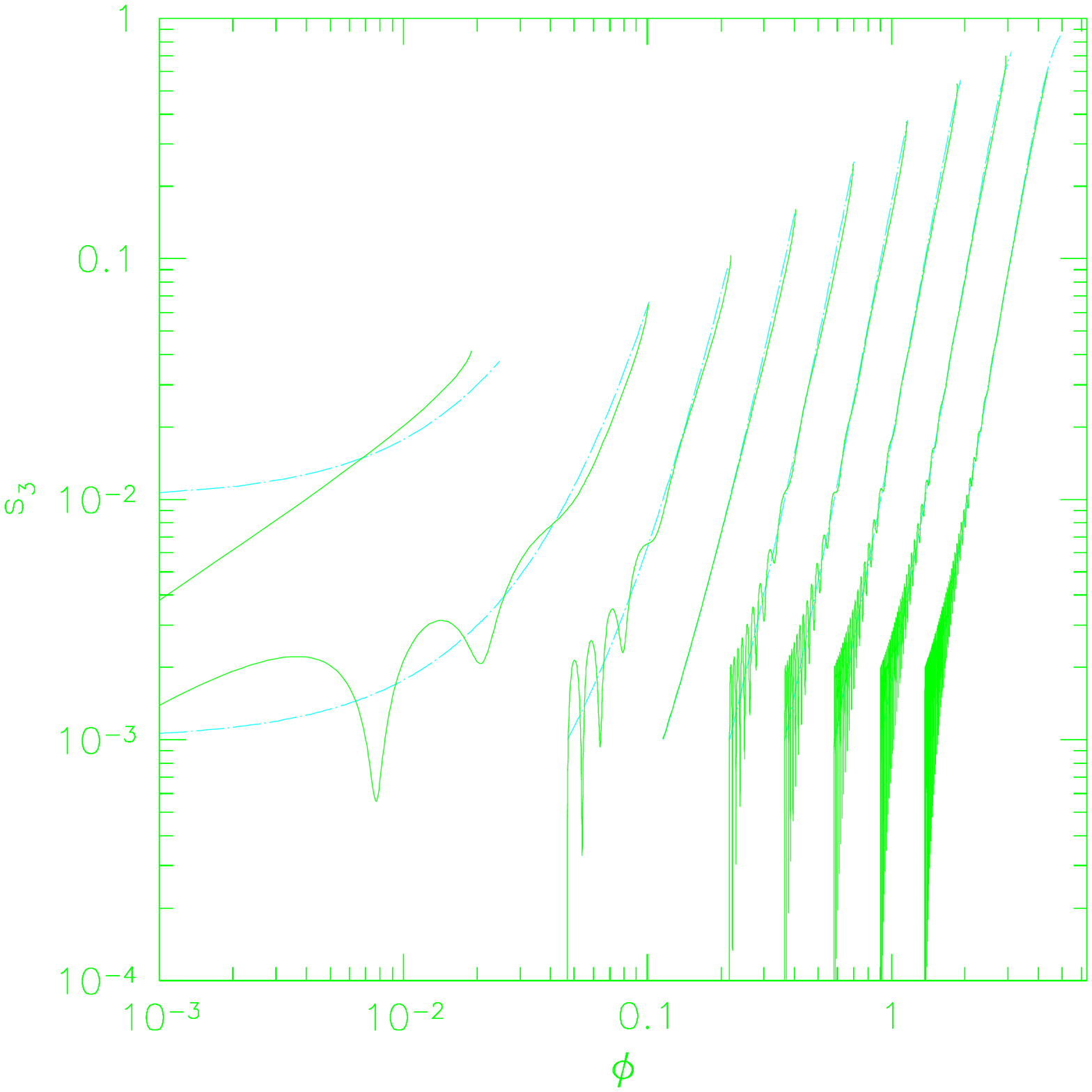}
\caption{Numerical solution and analytic approximation.  The
figures depict the exact numerical solution (solid lines) and the 
analytic solution based on the adiabatic approximation (dashed lines).
}
\label{fig:numersimp}
\end{figure}

We can also build an adiabatic approximation onto the exact solution
of \eqref{sevolrot} by allowing the magnitude of $\hatom$ to vary.  We take 
a wave polarized parallel to the initial value of $\hatom$.   As
$\hatom$ rotates about $\Upsilon$ and decreases in magnitude, 
${\bf s}$ rotates about $\hatom_\rmscr{eff}$ and follows 
the instantaneous direction of $\hatom_\rmscr{eff}$ as long as
\be
\left |\hatom \left ( \frac{1}{|\hatom|} \pp{ |\hatom |}{x_3} \right )^{-1}
\right | \gg 1.
\label{eq:adiarat}
\ee
Numerical integration of \eqref{sevol} for $|\hatom| = A x_3^{-6}$
and ${\bf \Upsilon} = - 1/10 {\hat x}_3$ ($\hatom$ rotates by one radian 
after the photon has travelled 10 units of distance) 
shows that the polarization follows the analytic solution described in 
the previous paragraph for
\be
\left |\hatom \left ( \frac{1}{|\hatom|} \pp{ |\hatom |}{x_3} \right )^{-1}
\right | \gtrsim 0.5
\label{eq:couplingratio}
\ee
and then freezes for values of $A$ ranging from 10 to $10^8$.
\figref{numersimp} depicts both the numerical results and the analytic
approximation.

Generically, if the polarization modes of the medium are linear, and
the wave begins in one of the modes, the final direction of the
polarization projected into the $S_1-S_2-$plane will depend on {\em
how much} the field direction has changed by the time of mode coupling.
The circular component of the polarization depends on {\em how fast}
the field direction is changing at the time of mode coupling.

\begin{figure}
\plotone{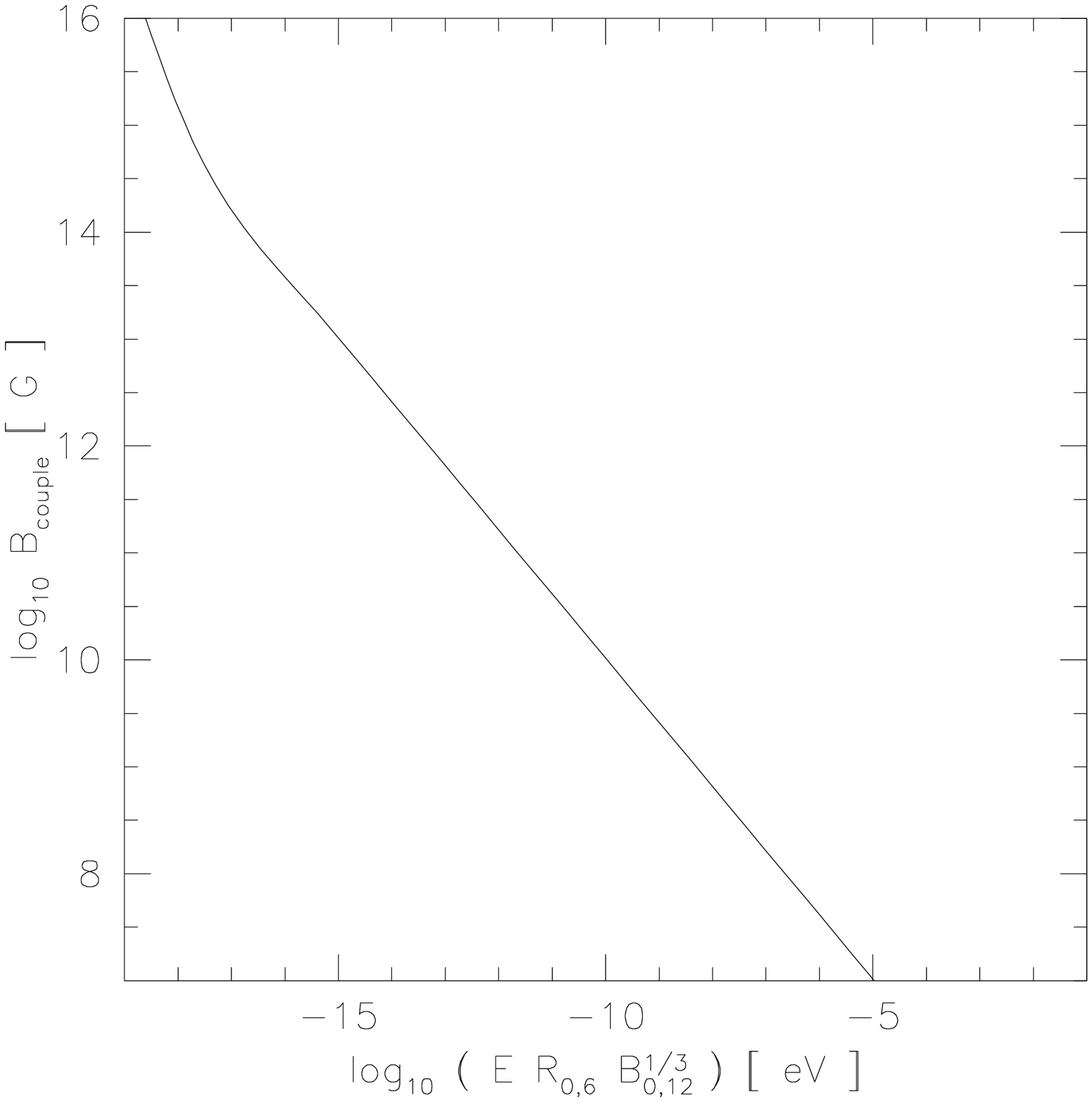}
\caption{The dipole magnetic field strength at coupling.  The strength of
the magnetic field at coupling depends on the magnetic moment of the
star and the energy of the photon.  The modes of higher energy photons 
couple later (further from the star).}
\label{fig:couple}
\end{figure}

By applying these results to a magnetic dipole in vacua, we find that an
outgoing photon's polarization will follow the analytic solution until
\be
\frac{3}{2} \dd{\ln \Delta n}{\ln B} \frac{B^{1/3}}{\Delta n} = k_0
R_0 B_0^{1/3},
\ee
after which its polarization in the observer's system freezes.  The
left side of the equality describes the environment at the point when
the polarization freezes; the right side depends on the point of
emission of the photon.
\figref{couple} shows the magnetic field
strength at coupling as a function of photon energy for typical
pulsar parameters.

If the modes begin to mix where $B\ll B_\rmscr{QED}$ (this is appropriate
for all but the lowest energy photons near the most strongly
magnetized neutron stars), we obtain
\ba
 \frac{ B_\rmscr{couple} }{B_0} &=& \left [ \frac{\alpha}{4\pi}
 \frac{2}{45} \left (\frac{B_0}{B_\rmmat{QED}} \right )^2 k_0 R_0
 \right ]^{-5/3} \\ &=& 1.95 \times 10^{-5} \left (
 \frac{E}{1~\rmmat{eV}} \right )^{-5/3} \left (
 \frac{B_0}{10^{12}~\rmmat{G}} \right )^{-10/3} \left
 (\frac{R_0}{10^6~\rmmat{cm}} \right )^{-5/3}.
\label{eq:ratio}
\ea
If this ratio is less than unity for a photon, the photon will travel
with its polarization modes decoupled during some
portion of its journey away from the star.  That is, if
\be
 E > 1.49 \times 10^{-3}~\rmmat{eV} \left (
 \frac{B_0}{10^{12}~\rmmat{G}} \right )^{-2} \left
 (\frac{R_0}{10^6~\rmmat{cm}} \right )^{-1},
\ee
the vacuum will decouple the polarization modes.
For parameters typical to neutron stars, the polarization modes
for all photons with $\lambda
\lesssim 800~\mu$m will be decoupled at least as
they begin their journey from the vicinity of the neutron star.
Although for photons of such low energy, the mode coupling induced by 
the plasma is likely to dominate (\eg \cite{Chen79}).

These effects may also be important for photons travelling through the
magnetospheres of strongly magnetized white dwarfs.  The most strongly
magnetized white dwarfs have $B \sim 10^9$~G and $R_0 \sim 10^9$~cm;
in the magnetospheres of these stars, the polarization modes of photons
more energetic than 1.5~eV will be decoupled by the vacuum birefringence.

\section{Rotating Pulsar Magnetospheres}

The photons that we observe from neutron stars have to travel through
a portion of the neutron star's magnetosphere before reaching us.  If
they travel through a region where \eqref{adiarat} holds, the photon's
polarization relative the observer's axes will change as the magnetic
field direction changes.  If we examine a photon which leaves the
surface of the star, once it has travelled a distance of several radii
it will be travelling approximately radially.  If the magnetic moment
of the star remains fixed during the journey and the modes remained
decoupled, the polarization directions of all the photons travelling
in a particular direction and polarization mode will align with each
other.

The plasma in the vicinity of a neutron star is polarized by passing
waves of sufficiently low frequencies.  \jcite{Chen79} used this
effect to explain the high linear polarization of radio emission from
pulsars even though one would expect the polarization modes where
the radiation is produced to vary from place to place.  The same
effect will operate at high frequencies through the vacuum
polarization of QED.

High frequency photons travelling through a pulsar magnetosphere may
travel a large distance before the modes couple.  During this travel,
the magnetic field of the pulsar may rotate a significant amount,
thereby changing the polarization modes in the observer's frame as the
photon travels.  Since high energy photons travel further than less
energetic ones, the higher energy photons will have their
polarizations dragged further by the rotating magnetic field.

\subsection{Analytic Treatment}

If we assume that the rotation rate of the projection of the magnetic
field onto the plane transverse to a photon's propagation changes
sufficiently slowly, we can apply the results of
\S~\ref{sec:adiabatic} to calculate the final polarization of the
photon.  What is required is the direction of magnetic field at the
point of recoupling and its rate of change.

This analytic treatment will be restricted to photons travelling in
the radial direction.  For a significant phase lag to develop, the
polarization modes must remain decoupled until well away from the
star.  Here outgoing photons emitted from near the stellar surface
will have approximately radial trajectories.  Furthermore, at these
distances both gravitational light bending and magnetic lensing
(\cite{Shav98lens}) may be neglected.

The photon's trajectory will be characterized
by the angles $\xi$ and $\theta$ the longitude and colatitude where
the photon left the star.  We shall take $\varpi$, the phase of the
star's rotation, to be zero when the photon leaves the surface.  The
magnetic pole lies at zero longitude and colatitude $\Psi$.  As the
photon travels away from the neutron star, the phase $\varpi$
increases and the local magnetic field direction changes.  The
criterion for recoupling in the weak-field limit when travelling
through a dipole field is
\be
 \frac{2}{15} \frac{\alpha}{4\pi} \left (
 \frac{B_{0,\rmscr{equator}}}{B_\rmscr{QED}} \right )^2 R_0^6 k_0
 \frac{\sin^2 \alpha}{r^6} \left | \dd{\sin^2\alpha}{r}
 \frac{1}{\sin^2\alpha} - \frac{6}{r} \right|^{-1} = \frac{1}{2}
 \label{eq:pulsar_recouple}
\ee
In this equation, the angle $\alpha$ designates the magnetic colatitude
of the photon which changes as the star rotates underneath it.  It
satisfies the following expression
\be
 \cos \alpha = \sin \theta \cos ( \xi + \varpi ) \sin \Psi + \cos \Psi
 \cos \theta. 
\ee
The second important angle is the angle between the observer's axes
and the local magnetic field direction in the plane of the sky.  For a
magnetic dipole, the projection of the local magnetic direction and
the projection of the magnetic moment ($\hat{\bf m}$) onto the
tangential plane are colinear.  Unless the line of sight is directed
down the rotation axis, we can use the projection of the rotation axis
($\hat{\bf z}$) as one of the reference axes for measuring the
polarization of the radiation.  These projected vectors are defined by
\ba
{\bf z}_\rmscr{p} &=& {\hat{\bf z}} - {\hat {\bf r}} \cos \theta  \\
{\bf m}_\rmscr{p} &=& {\hat{\bf m}} - {\hat {\bf r}} \cos \alpha. 
\ea
The travelling photon does not distinguish between a magnetic field
pointing in one direction or in the exactly opposite direction (see
\eqref{omegaqeddef}); therefore, we only need to know the angle
between ${\bf z}_\rmscr{p}$ and ${\bf m}_\rmscr{p}$ modulo $\pi$.
Calculating the cross product suffices to give the magnitude of the
angle,
\be
\left ( {\bf z}_\rmscr{p} \times {\bf m}_\rmscr{p} \right ) \cdot 
{\hat {\bf r}} = z_\rmscr{p} m_\rmscr{p} \sin \phi,
\ee
which yields
\be
\tan \phi = 
\frac{\sin (\xi + \varpi)}{\cos\theta\cos(\xi+\varpi)-\cot\Psi\sin\theta}
\label{eq:pulsar_sinphi}.
\ee
${\bf \Upsilon}$ in this case is given by
\be
{\bf \Upsilon} = -2 \dd{\phi}{\varpi} \frac{2\pi}{cP} ~{\hat 3}.
\label{eq:pulsar_upsilon}
\ee
So the final polarization of the photon in the observers frame is
given by \eqref{omegaeff} evaluated at the moment of recoupling.

The calculation of the polarization evolution in the analytic
treatment proceeds as follows,
\begin{enumerate}
\item
Choose the colatitude of the observer ($\theta$), the longitude
where the photon is emitted ($\xi$), a reference frequency and
magnetic field strength.
\item 
Given the period of the pulsar ($P$), a photon's radius is given by 
\be
r = \frac{c P}{2\pi} \varpi + R_0.
\label{eq:radius_phase}
\ee
For a given value of $\varpi$, the phase, solve for
$B_{0,\rmscr{equator}}^2 k_0$ assuming that
\eqref{pulsar_recouple} is satisfied, and calculate the magnitude
of $\hatom$ at recoupling.
\item
Substitute this value of $\varpi$ into \eqref{pulsar_sinphi} to
calculate the direction of $\hatom$ at recoupling.
\item
Finally, calculate ${\bf \Upsilon}$ at recoupling using
\eqref{pulsar_upsilon}.  The final polarization according to this
analytic adiabatic approximation lies along
\be
\hatom_\rmscr{eff} = \hatom - {\bf \Upsilon}.
\ee
\item
This procedure can be repeated for other values of the longitude
($\xi$) to generate a light curve.
\end{enumerate}

In the case where the line of sight is aligned with the rotation axis,
many of these geometric considerations vanish, and the problem reduces
identically to that treated in \S~\ref{sec:adiabatic}.
\begin{figure}
\plotone{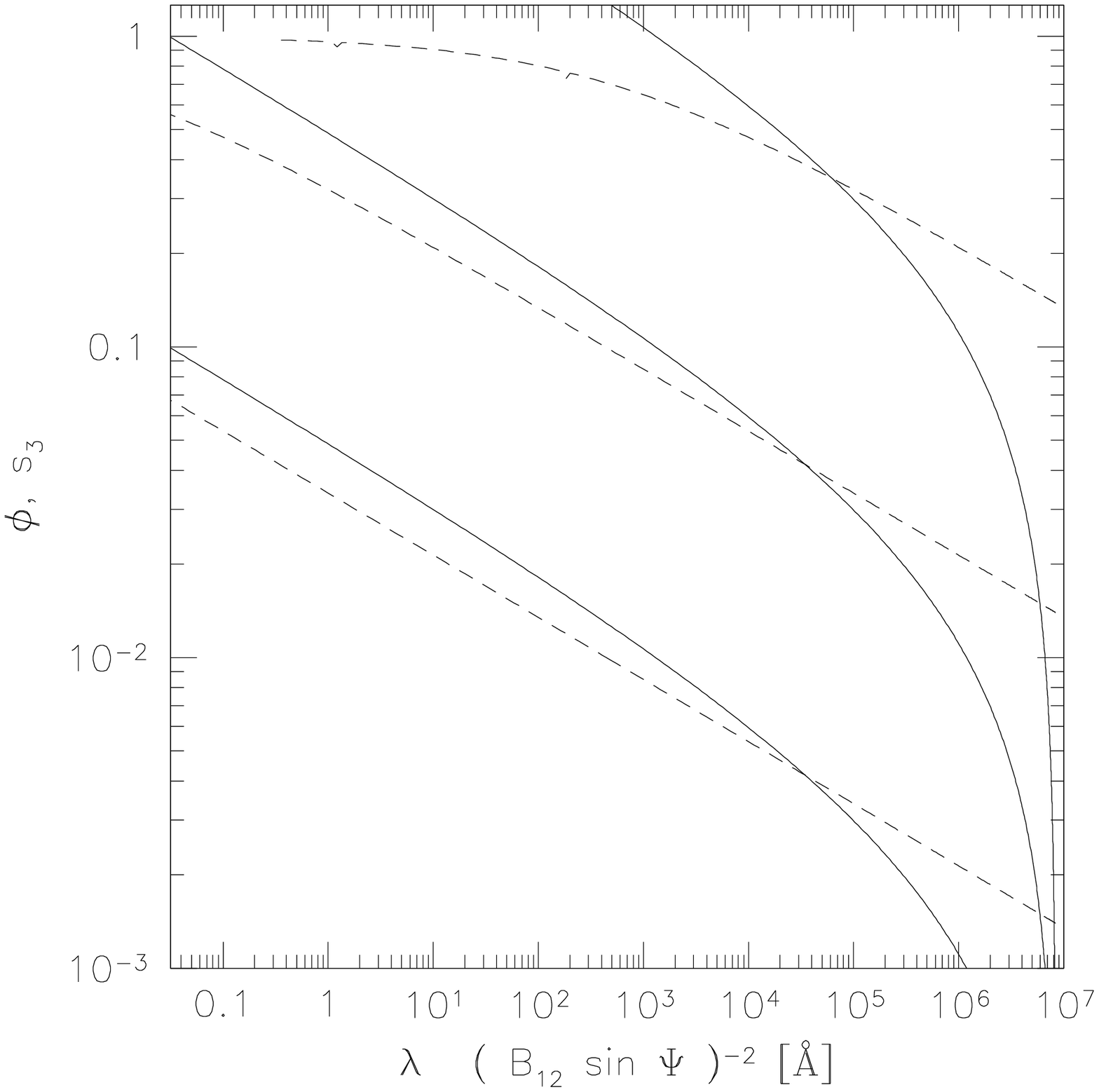}
\caption{The observed polarization as a function of frequency for a
line of sight parallel to the rotation axis in the presense of a
rotating dipole field.  The panel depicts the polarization angle
(solid line) and circular component (dashed line, normalized to $|{\bf
s}|$) as a function of wavelength.  The lines from top to bottom trace
the solutions for $P=1, 10,$ and $100$~ms respectively.}
\label{fig:resana_0}
\end{figure}

In this situation we can derive expressions for both the final position 
angle,
\ba
\phi &=& \frac{2 \pi}{c P} \left \{ \frac{1}{90}
\left [ \alpha \frac{90^4}{\pi} k_0 R_0^6
\left ( \sin \Psi \frac{B_{0,\rmscr{equator}}}{B_\rmscr{QED}} \right
)^2
\right]^{1/5} - R_0 \right \} 
\label{eq:phiaxis}
\\
   &\approx& 7.7 \times 10^{-4} \left ( \sin \Psi
\frac{B_{0,\rmscr{equator}}}{10^{12}~\rmmat{G}}  \right )^{2/5}
\left ( \frac{E_\rmscr{photon}}{1~\rmmat{eV}} \right )^{1/5}
\left ( \frac{P}{1~\rmmat{sec}} \right )^{-1}
\left ( \frac{R_0}{10^6~\rmmat{cm}} \right )^{6/5}
\ea
and the circular component,
\be
s_3 / |{\bf s}|= \frac{4 \pi}{c P} \left [ \left ( \frac{3}{r} \right )^2 +
\left ( \frac{4\pi}{c P} \right )^2 \right ]^{-1/2} ,\\
\ee
where
\ba
r = \phi \frac{c P}{2 \pi} + R_0.
\ea
If $|s_3| / |{\bf s}| \ll 1$, the following approximation holds,
\be
s_3 / |{\bf s}| \approx \frac{2}{3} \phi + \frac{4}{3}\pi \frac{R_0}{c P}.
\ee

For a more general pulsar geometry we can still use \eqref{phiaxis}
to characterize how radiation at various frequencies is polarized.
Specifically, the position angle can be converted to a time lead for
the polarization angle of high-frequency relative to low-frequency
radiation.  This time lead given by $P \phi/(2\pi)$ is
independent of the period of the pulsar.
\begin{figure}
\plotone{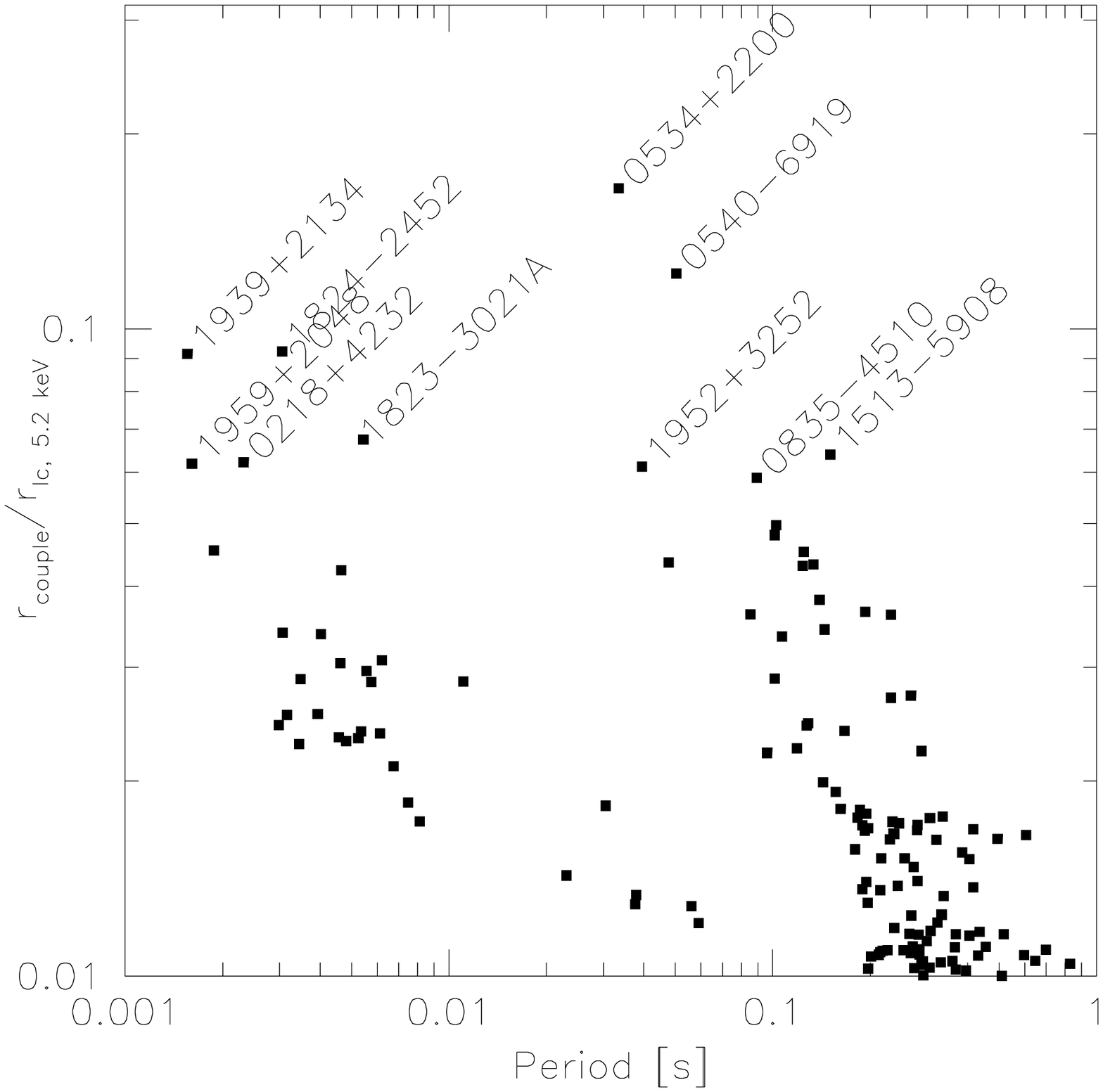}
\caption{The ratio of the coupling radius at 5.2~keV to the radius of
the light cylinder for the known radio pulsars. Only pulsars with
$r_\rmscr{coupling}/r_\rmscr{lc}>0.01$ are plotted.}
\label{fig:pulsarcat}
\end{figure}
\figref{pulsarcat} plots the known radio pulsars (\cite{Tayl93}) and
the approximate phase lead ($r_\rmscr{couple}/r_\rmscr{lc}$)
expected at 5.2~keV relative to zero-energy photons.  If
$r_\rmscr{couple}$ is a large fraction of the radius of the light
cylinder, it may be possible to study the geometry of the magnetic
field at large distances from the pulsar.  For the Crab pulsar the
radius at coupling for 5.2~keV is about 16\% of the light-cylinder
radius.  The millisecond pulsars, PSR~J1939+2134 and PSR~J1824-2452,
have $r_\rmscr{couple}$ of over 9\% of $r_\rmscr{lc}$ at 5.2 keV.  

\subsection{Numerical Treatment}

The analytic techniques outlined in the previous subsection provide
insights to interpret and extend the results from a direct numerical
integration of \eqref{sevol}. Nevertheless, a numerical integration is
unavoidable if we wish to calculate the phase shifts under general
conditions -- when the observer's inclination angle is nontrivial and
the field is not strictly dipolar.

The numerical integration assumes two basic assumptions:
\begin{enumerate}
 \item We assume that either
 \begin{enumerate}
 \item The magnetosphere is co-rotating with the NS.  This implies that
 the local magnetic field is at any given moment aligned with the NS,
 irrespective of the distance. (A deviation from this behavior will
 indicate where the co-rotating magnetosphere ends), {\it or}
 \item The magnetic field is given by that of a magnetized conducting 
	sphere rotating {\it in vacua} (\cite{Deut55}, \cite{Barn86})
 \end{enumerate}
 \item The photon is traveling radially. That is to say, the emission
 process takes place well within the region where the polarization
 states couple.
\end{enumerate}
With these assumptions considered, the integration is achieved in a
straightforward manner. A photon is followed from an initial radius
$R_0$ to a radius $r_\rmscr{final}$ that is beyond the recoupling 
distance using the equation:
\be
{\partial {\bf s} \over \partial r} = \hat{\bf \Omega} \times {\bf s}.
\ee
Each radial step is integrated forward using the fourth order
Runge-Kutta algorithm. To do so, one has to calculate the birefringent
vector $\hat{\bf \Omega}$ using \eqref{pulsar_sinphi} for the angle of
the magnetic field and \eqref{delta_n_weak} for the strength of the
birefringence.  To calculate this vector for the \jcite{Deut55} fields, we 
use Eq. 7 of \jcite{Barn86}.
The rotation phase of the star is related to the radial
coordinate through \eqref{radius_phase}.

\subsubsection{The Crab pulsar}
\begin{figure}
\plotone{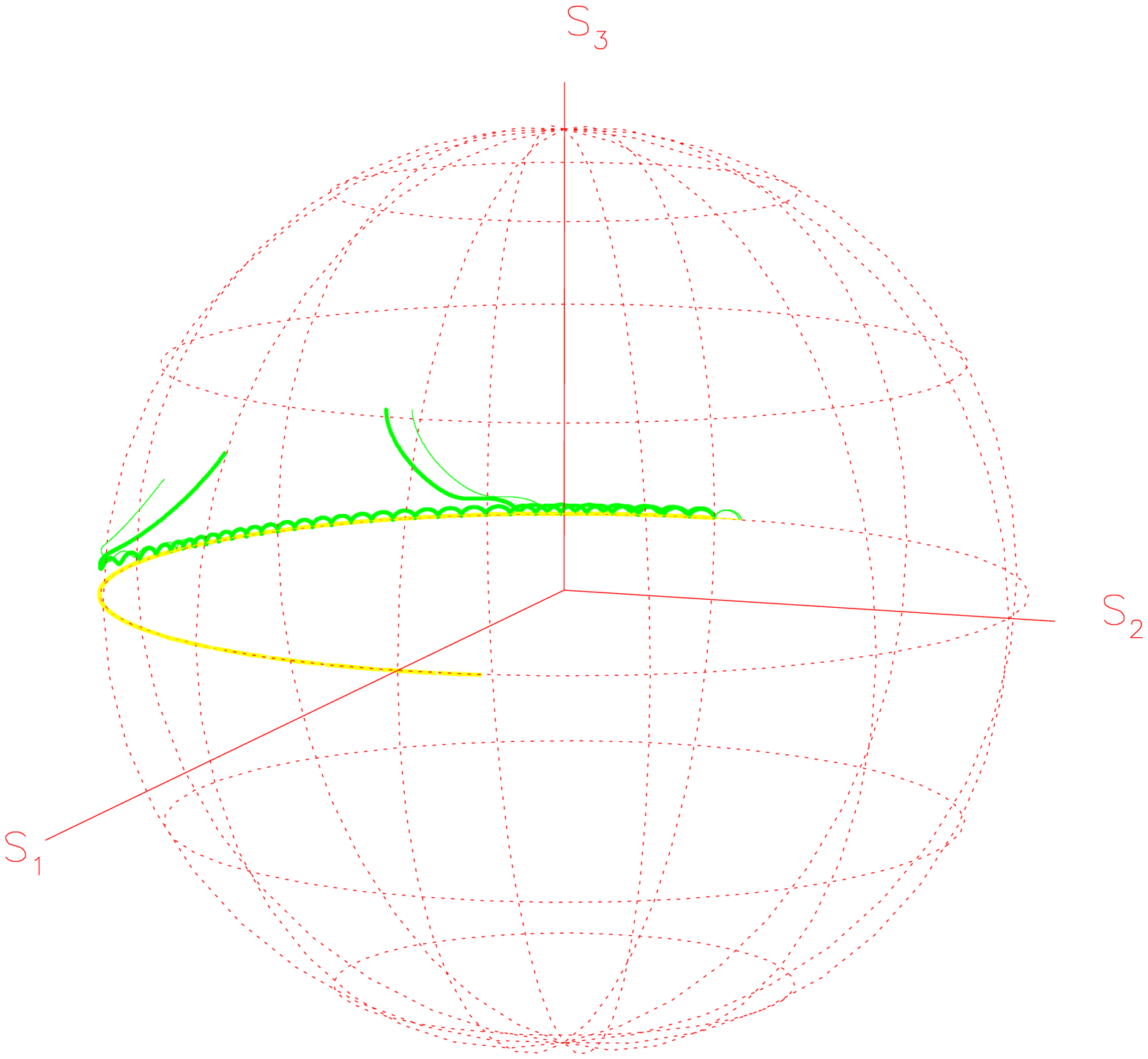}
 \caption{The evolution of the Stokes parameters on the Poincar\'e
 sphere for photons leaving the surface of the Crab pulsar at the
 moment the observer crosses the Prime Meridian.  The bold lines
 depict the corotating dipole, and the light ones the Deutsch field.
 The energies plotted are 5.2~keV and 5.2~MeV. The bold line in the
 1-2 plane is the birefringent vector $\hatom$. The photons'
 polarization vectors and the birefringent vector all start from the
 -1 direction. The lower energy photons follow $\hatom$ 
 for a shorter time than do the higher energy ones. Since the mode 
 recoupling is not instantaneous, a circular component ($S_3$) is generated. }
\label{fig:stokes_evol}
\end{figure}

\begin{figure}
\plottwo{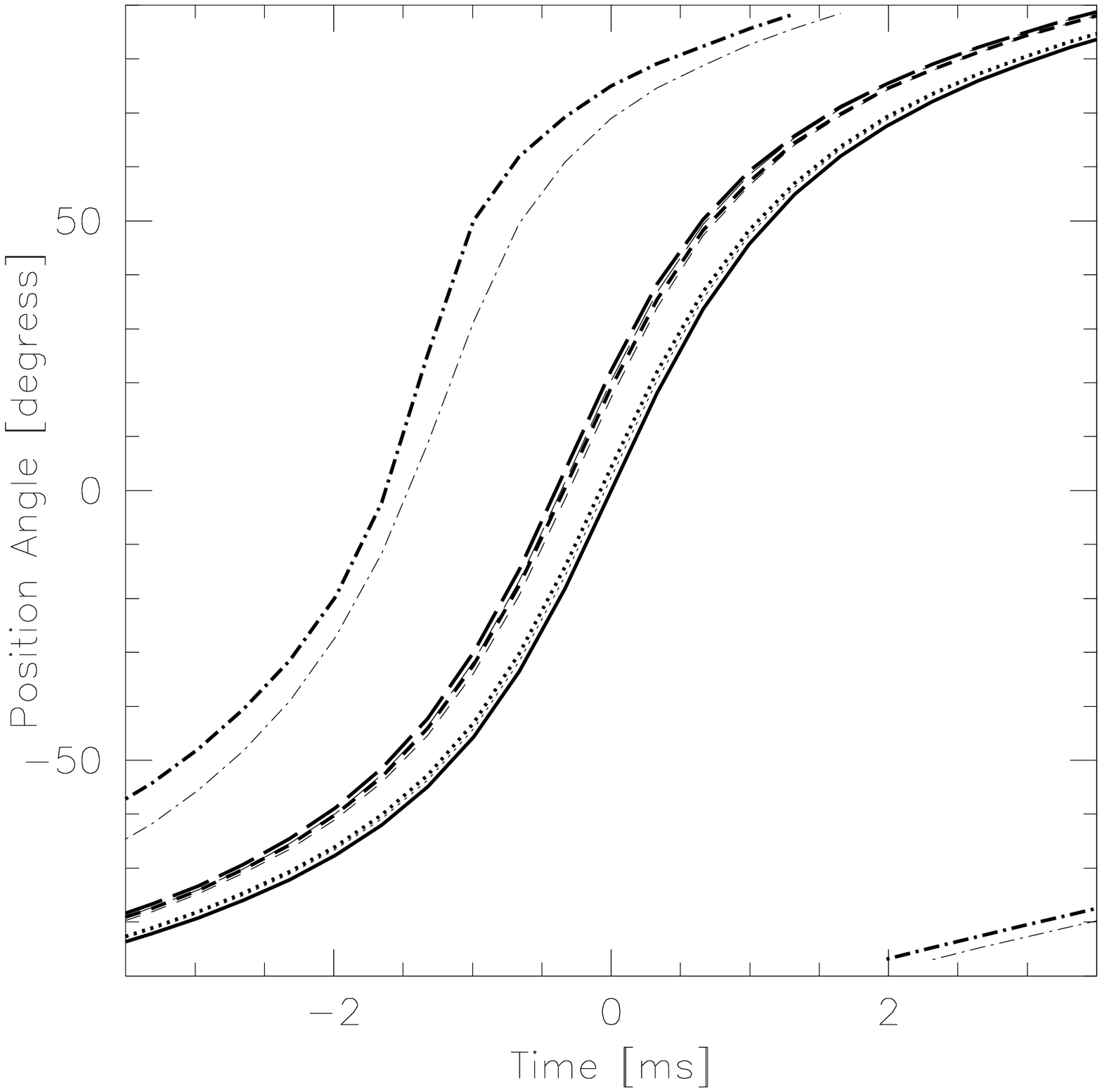}{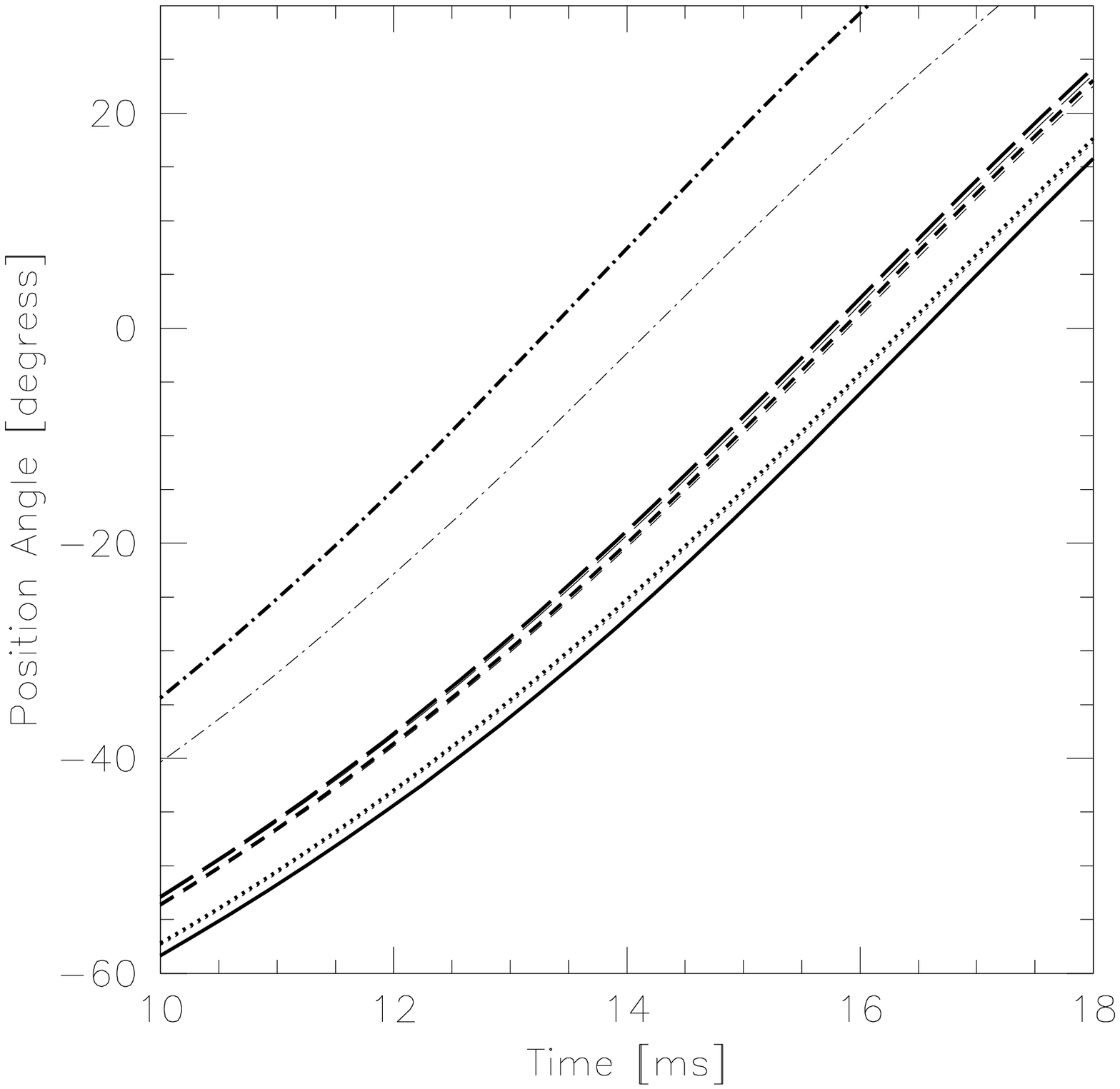}
\caption{The position angle as a function of photon energy and
time for our model of the Crab pulsar.  From left to right, the
energies are 5.2~Mev (dot-dashed), 5.2~keV (long-dashed),
2.6~keV (short-dashed), 4.48~eV (dotted) and zero energy (solid).  The
bold lines follow
the corotating dipole model, and the light lines follow the Deutsch 
model.}
\label{fig:crab}
\end{figure}

The Crab pulsar is among the most thoroughly studied astronomical
objects.  Its fast period and moderately strong magnetic field make it
an ideal example for this process.  We take $\theta=54^\circ$ and
$\Psi=64^\circ$ which reproduces the magnitude of the polarization
swing for the pulse and interpulse (\cite{Smit88}),
and the field strength at the
magnetic equator to be $1.9\times 10^{12}$~G (\cite{Tayl93}).
The results of this
calculation are depicted in \figref{stokes_evol} and \figref{crab}.
If the field is indeed a corotating dipole, the delayed coupling
of the polarization modes results in the polarization
of high frequency radiation leading that of lower energies.
The lead time estimate from \eqref{phiaxis} works well away from the
pulse or interpulse where it slightly over or underestimates the lead
time.  However, if the neutron star is surrounded by a Deutsch field,
this lead is retarded slightly.

\begin{figure}
\plotone{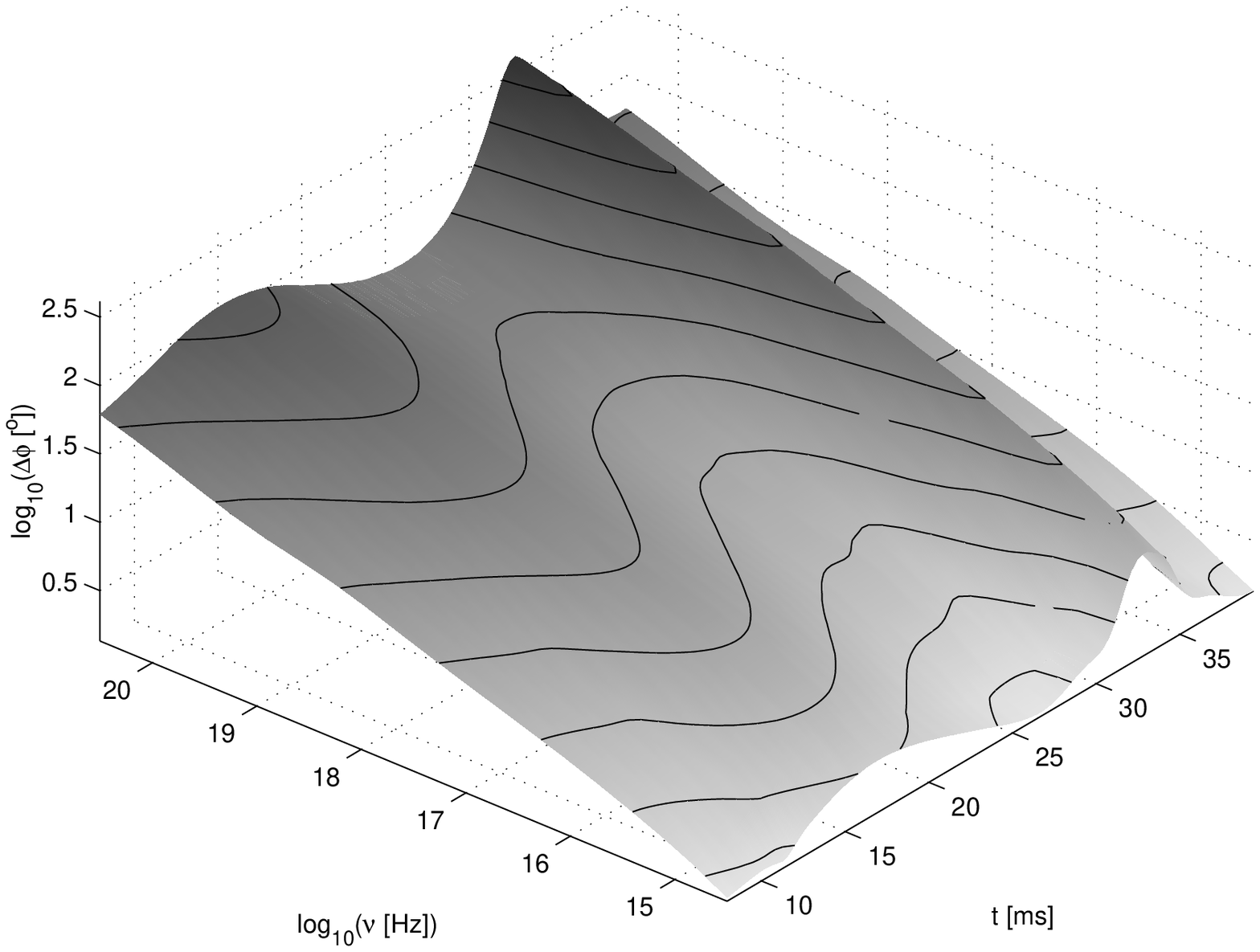}
\caption{The difference between the final position angle of a photon emitted
at a particular time and energy and the position angle of a zero
energy photon emitted at the same time.  The model is for the Crab
pulsar surrounded with a Deutsch field.}
\label{fig:crab3d}
\end{figure}

Since the coupling of the modes occurs gradually, a significant
circular component will develop when the modes are weakly coupled.
If the radiation initially has a circular component, this initial 
component is washed out when one measures the circular
polarization over a finite bandpass, leaving only the value of $s_3$ 
produced by the coupling.
\begin{figure}
\plotone{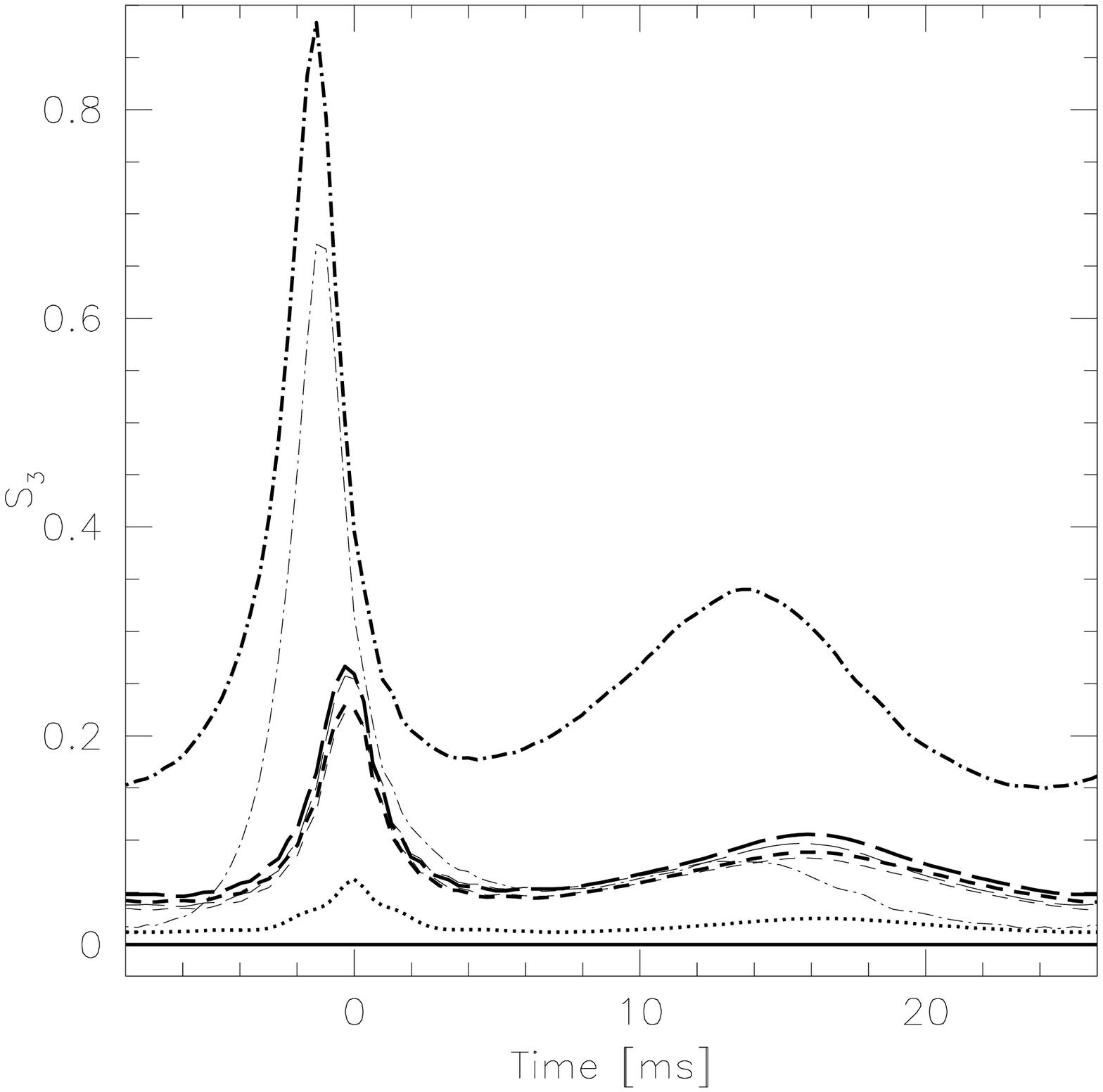}
\caption{The circular component of the radiation as a function of
photon energy and time for the Crab pulsar, assuming the initial
polarization is complete (i.e., $|{\bf s}|=1$; otherwise, it should be
normalized to its smaller value).  The energies are 5.2~MeV
(dot-dashed), 5.2~keV (long-dashed), 2.6~keV (short-dashed), 
4.48~eV (dotted) and zero
energy (solid).  The bold lines follow the corotating dipole model,
and the light lines follow the Deutsch model.}
\end{figure}
Both the Deutsch fields and the corotating dipole fields result in a
significant amount of circular polarization (up to 30\% of the total
polarization for the corotating dipole at 5.2~keV or 7\% of it at
4.5~eV).  The circular polarization generated by the rotating Deutsch
field peaks at about 65\% of the total polarization at the highest
frequency studied.

\subsubsection{RX~J0720.4-3125}

\begin{figure}
\plottwo{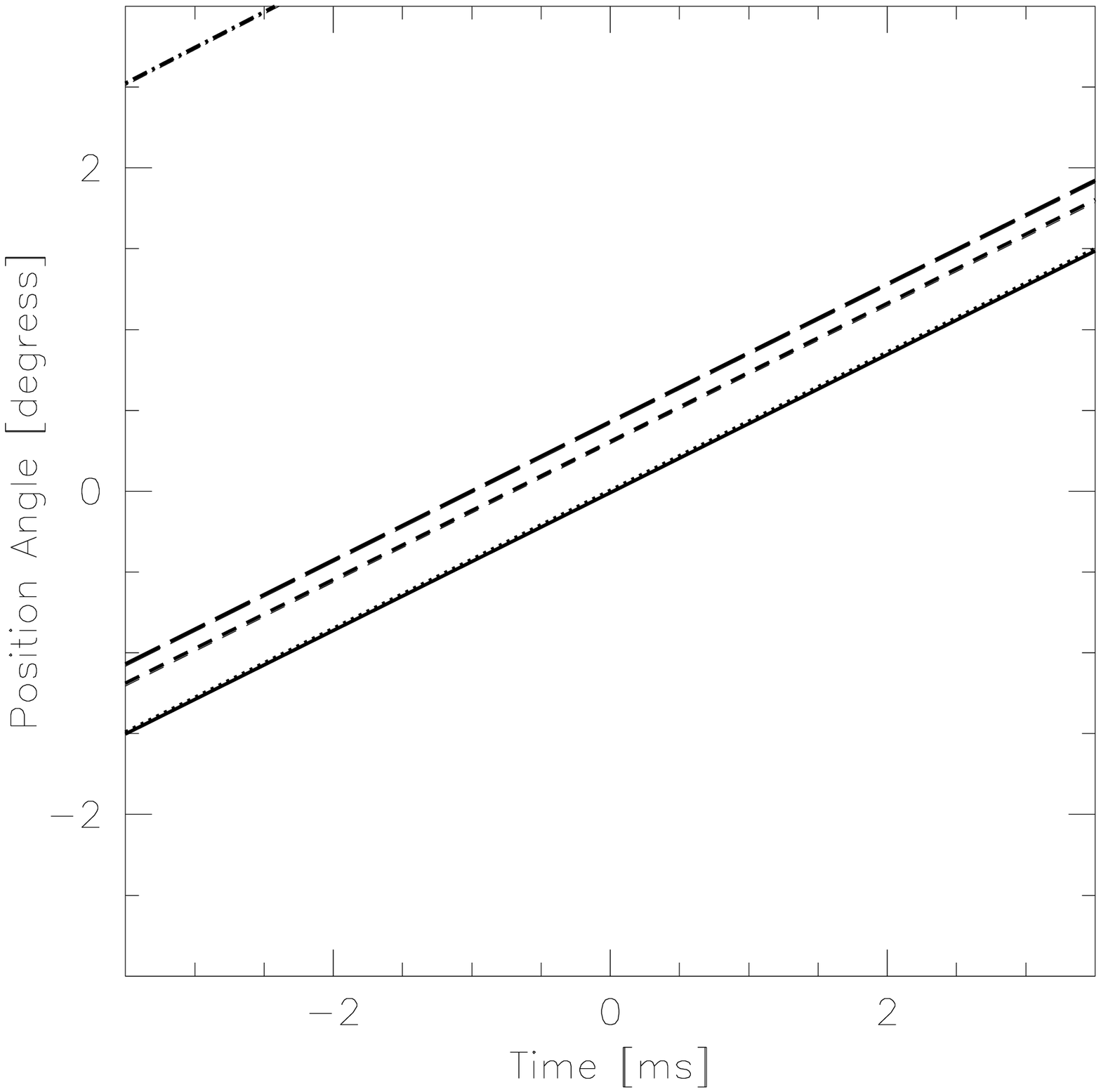}{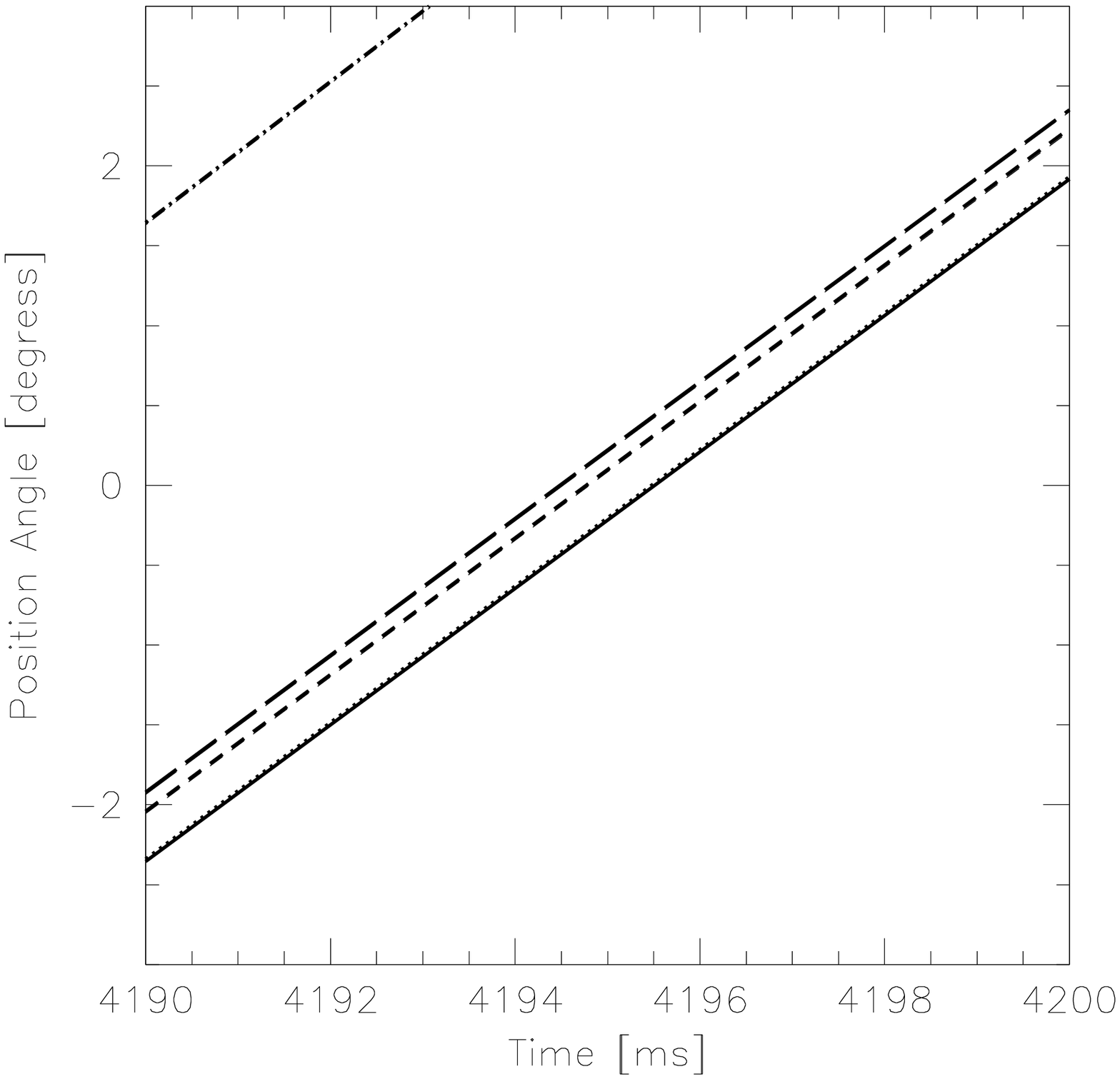}
\caption{The position angle as a function of photon energy and
time for our model of RXJ~0720.4-3125.  We have taken
$B_{0,\rmscr{equator}}=10^{14}$~G, $\theta=85^\circ$ and
$\Psi=90^\circ$.  From left to right, the
energies are 5.2~Mev, 5.2~keV, 2.6~keV, 4.48~eV and zero energy.}
\label{fig:rxj}
\end{figure}

RX~J0720.4-3125 is an isolated neutron star candidate suspected to be
close to the Earth.  \jcite{Heyl98rxj} from cooling arguments and its
current period of 8.391~s estimate that it has a magnetic field of
$B\sim 10^{14}$~G.  From \eqref{phiaxis} we estimate the time lead for
a given frequency to be approximately five times larger than for the
Crab pulsar.  As we see from \figref{rxj}, since the period of 
RX~J0720.4-3125 is much larger than that of the Crab, the expected
change in position angle over the larger lead time is very small.
Furthermore, since the coupling of the modes occurs well within the
light cylinder, the difference between the Deutsch and the corotating
dipole model is negligible.
Since the circular polarization that develops during coupling is
proportional to the rotation frequency of the pulsar, very little
circular polarization is evident in the emergent radiation -- $S_3$
peaks at 2.7\% of the total polarization, coincident with the pulse
and interpulse.

\begin{figure}
\plotone{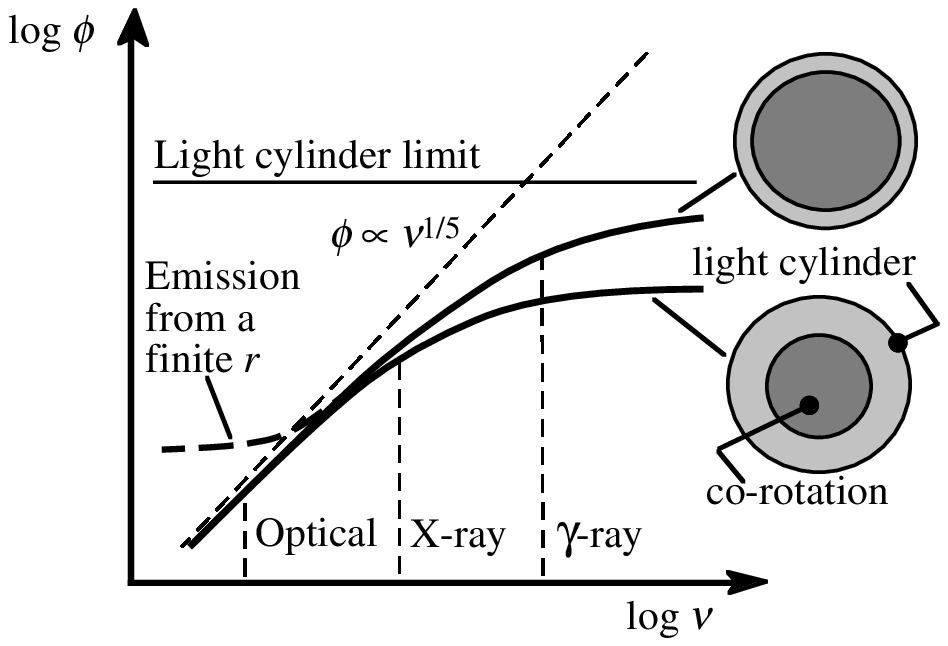}
 \caption{A schematic depiction of the expected phase lag at a given
 rotation phase of the NS.  Lower frequency photons follow the
 changing magnetic field only over a short distance from the neutron
 star, and thus the modes couple when the local direction of the
 magnetic field has not changed much since emission.  The polarization 
 modes of higher energy photons couple further away and thus follow
 its changing direction further, creating a phase lead between the
 photon and the NS dipole direction.  The magnetosphere rotates as a
 solid body only up to the co-rotation radius (which must be less than
 the radius of the light cylinder); consequently, a photon's phase
 lead will not be increased anymore once it begins traveling through
 the retarded magnetic field. If the emission process takes place at a
 finite radius, the lower energy photons could already be emitted
 outside the decoupled region. In such a case, they would suffer no
 phase lag.}
\label{fig:phase_saturation}
\end{figure}

\section{Discussion}

The vacuum polarization induced by quantum electrodynamics decouples
the polarization modes in the strong magnetic field surrounding
neutron stars.  As radiation travels away from the neutron star, the
modes of low-frequency radiation couple first; consequently, the
position angle of low frequency emission will lag behind that of
higher frequency radiation emitted coincidently and in the same mode.
Puslars often exhibit mode switching as a function of phase and
frequency.  An inelegant way to avoid confusion is to examine the
position angle modulo $\pi/2$.
By measuring the position angle as a function of frequency, one can
determine not only the location of the emission but also probe the
structure of the magnetic field near the light cylinder using
sufficiently high frequency radiation.  

Previous authors have calculated a similar effect where the plasma
decouples the modes at radio frequencies, assuming that the coupling
occurs instantly (\cite{Chen79}; \cite{Barn86}).  To relax this
assumption, we use the equations describing the propagation of 
polarized radiation through a polarized and magnetized medium, under
the assumptions of geometric optics.  These results extend
those of \jcite{Kubo81} to the case of a magnetized and polarized
medium.  Linearly polarized radiation will develop a
circular component during the gradual coupling of the modes.  

In the case of QED in a vacuum, the propagating modes are purely linear;
therefore, the decoupling will wash out any initial circular component
averaged over a finite bandpass.  The only circular component present
in the outgoing radiation is produced during coupling and measures 
the ratio of the coupling radius to the radius of the light cylinder.
This circular component provides an independent validation of the
effect.

Since high frequency radiation is generally measured incoherently,
determining the Stokes parameters becomes more difficult with
increasing photon energy.  Blueward of the ultraviolet, measuring
circular polarization is impractical.  The Spectrum-X-Gamma mission
will carry a stellar x-ray polarimeter (SXRP) which will be especially
sensitive at 2.6~keV and 5.2~keV, the first and second-order Bragg
reflections off of graphite.  Future instruments may be able 
to measure both linear and circular polarization at yet higher
energies.  

From a theoretical point of view, the study of the propagating modes
of a strongly magnetized vacuum is simpler than analysing those of a
strongly magnetized plasma, especially when the properties of the
plasma are not well known.  However, given the magnetic permeability
and the dielectric permittivity of the circumpulsar plasma, these
results can be applied to study plasma-induced decoupling on the
properties of polarized radio emission from pulsars.  We leave the
details of this process for a subsequent paper; however, our general
arguments apply here as well.

Both the plasma-induced and vacuum-induced decoupling can be used to
probe the structure of pulsar magnetospheres.  A measurement of the
circular polarization produced by the mode coupling yields an estimate
of the radius of the coupling.  The frequency corresponding to a given
degree of circular polarization measures the radial gradient of third
power of the difference between the indices of refraction of the two
modes.  This results from the approximate inequality
\eqref{couplingratio}.  Like the arguments used in
\S~\ref{sec:adiabatic}, it is generic to waves travelling through a
birefringent medium in the limit of geometric optics.

Unlike the phase lead effect found by previous authors for the case
of the plasma alone, the circular polarization provides a local probe
of the pulsar magnetosphere without having to compare observations
over a wide range of energy.  As long as the radiation is produced in a
region where the modes are decoupled, the interpretation of the
circular polarization is straightforward.

Measurements of the position angle of the radiation as a function of
phase and frequency can also yield the same information, since the
induced circular polarization is simply related to the phase lead.
Furthermore, measuring all four Stokes parameters can elucidate the
location of the emission process since below a critical frequency the
modes will be coupled at the point of emission and the observed
position angle will be constant with frequency.  The circular
polarization of the first modes which do travel through a
decoupled region yields the radius of the emission.  The details of
the interpretation will only depend weakly on the medium which
decouples the modes -- plasma or vacuum.

We have derived equations which describe the propagation of polarized
radiation through a magnetized and polarized medium.  By applying
these results to the strongly magnetized regions surrounding neutron
stars, we have found several observational probes of pulsar
magnetospheres and emission.

\bibliographystyle{jer}
\bibliography{mine,physics,ns,qed,glitch}
\end{document}